# FuGeIDS: Fuzzy Genetic paradigms in Intrusion Detection Systems

**Rajdeep Borgohain**
Department of Computer Science and Engineering, Dibrugarh University Institute of Engineering and Technology,
Dibrugarh, Assam
Email: Rajdeepgohain@gmail.com

------------------------------------------------------------------ABSTRACT------------------------------------------------------------------
With the increase in the number of security threats, Intrusion Detection Systems have evolved as a significant countermeasure against these threats. And as such, the topic of Intrusion Detection Systems has become one of the most prominent research topics in recent years. This paper gives an overview of the Intrusion Detection System and looks at two major machine learning paradigms used in Intrusion Detection System, Genetic Algorithms and Fuzzy Logic and how to apply them for intrusion detection.

## 1. INTRODUCTION

With the leap in information flow across the internet, the network scale is expanding on a daily basis, and with it is increasing the threats to the systems connected to the networks. An attack on important and confidential data is one of the foremost concerns for the users. These attacks may come in various forms including hacks, virus, worms, steganography [3, 26], dictionary attacks [5, 40, 11] and denial of service attacks [36, 37] etc. With these threats looming every time a user goes online, it is important to have a mechanism that could analyze these threats and defend the system against them. Many researchers have suggested numerous techniques to deal with security threats which include multifactor authentication [16], multipath routing [14, 33, 18, 31, 21], and biometric security measures [38, 6] etc. Besides adopting these measures, conventional measures like firewalls, antiviruses, network analyzers etc. are also used. Among these, one of the most significant counter measures [10, 29, 30, 19] against security threats is the Intrusion Detection System also known as IDS.

The basic job of the Intrusion Detection System is to monitor the incoming and outgoing traffic from a system and looks for any abnormal behavior in the network activity which indicates a possibility of threat. The Intrusion Detection System with certain predefined rules alerts the system of suspicious patterns in the network activity.

One of the most significant aspects of Intrusion Detection System is the use of Artificial Intelligence [39] to train the Intrusion Detection System about the possible threats. The Intrusion Detection can gather information about the various traffic patterns and rules can be formed based on these patterns, to distinguish between normal traffic and anomalous traffic in the network.

Though many Artificial Intelligence techniques are being used in Intrusion Detection System [24], we look at two of the most prominent techniques, Fuzzy Logic and Genetic Algorithms. Intrusion Detection Systems can be developed based on Fuzzy logic and Genetic Algorithms separately or both the techniques can be combined to develop a Fuzzy-Genetic Intrusion Detection System [41, 42]. Genetic Algorithms use audited data from network to derive a set of classification rules and fitness function judges the quality of the rules, which are then applied in the real world environment to counter network intrusions [7]. As attacks on systems may not have a discrete pattern, fuzzy logic can be used to detect attack patterns which may have a behavioral pattern between normal and anomalous. Moreover, fuzzy logic also helps to lower the rate of false positive alarms [11].

The rest of the paper is organized in the following way, Section 2 gives an overview of the Intrusion Detection System. Section 3 discusses the different types of Intrusion Detection Systems. In Section 4, some of the artificial intelligence techniques used in IDS are discussed. Section 5 discusses the Intrusion Detection System using Fuzzy Logic. Section 6 discusses the Intrusion Detection System using Genetic Algorithms and at the end Section 7 gives a conclusion.

## 2. OVERVIEW OF INTRUSION DETECTION SYSTEM

Intrusion detection Systems are used to monitor unwanted and malicious network traffic and take appropriate action if such situation occurs. The Intrusion Detection System uses a sensor to collect the information about traffic data. The sensor is connected to the network line using a network tap. The traffic data collected in the sensor is then sent to the Intrusion Detection System collector, which analyses the traffic data sent from the sensor. The collector analyses the data using *Signature Detection technique* or the *Anomaly Detection technique* and term the data as either a threat or normal traffic data. The Intrusion Detection System Manager lets the user configure the Intrusion Detection System.

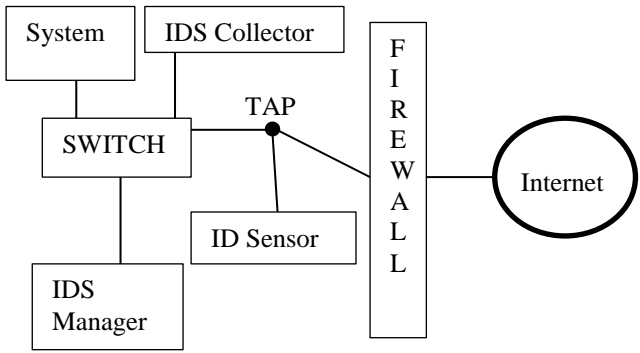

Figure 1. Intrusion Detection System

## 3. TYPES OF INTRUSION DETECTION SYSTEM

Intrusion Detection Systems can be classified based on their monitoring activity, detection technique and their response to the attack.

### 3.1 INTRUSION DETECTION BASED ON MONITORING ACTIVITY

The Intrusion Detection Systems can be broadly divided into two types based on the fact that whether they monitor the whole network or a particular host. Accordingly, they are termed *Host based Intrusion Detection System (HIDS)* and *Network based Intrusion Detection System (NIDS)*.

#### 3.1.1 HOST BASED INTRUSION DETECTION SYSTEM

The Host Based Intrusion Detection System resides in the host and traffic data is analyzed individually in each host. The IDS monitors the various file systems, network events and system calls to detect any possible threat to the system.

#### 3.1.2 NETWORK BASED INTRUSION DETECTION SYSTEM

The NIDS monitors the packets passing through the entire network and analyses the packets. Network Based Intrusion Detection System is particularly useful for monitoring traffic of many systems all at once.

### 3.2 INTRUSION DETECTION BASED ON DETECTION TECHNIQUE

Intrusion Detection System is also classified on the basis of the technique used by the IDS to look up for vulnerabilities. They are mainly classified into *Signature Based Detection* and *Anomaly Detection*.

#### 3.2.1 SIGNATURE BASED DETECTION

The Signature Based Detection compares a possible threat with the attack type already stored in the IDS. The limitation of this type of detection technique is that if any new type of threat comes which is not already known to the IDS, the system becomes vulnerable to that attack.

#### 3.2.2 ANOMALY BASED DETECTION

The anomaly based detection is a detection technique by which the IDS looks for vulnerabilities based on rules set forth by the user and not on the basis of signatures already stored in the IDS. This type of detection usually uses Artificial Intelligence to distinguish between normal traffic and anomalous traffic.

### 3.3 INTRUSION DETECTION BASED ON RESPONSE TO ATTACK

Depending on the response to a security threat, the Intrusion Detection Systems can be classified as *Active Intrusion Detection System* (also known as *Intrusion Prevention System*) and *Passive Intrusion Detection System.*

#### 3.3.1 ACTIVE INTRUSION DETECTION SYSTEM

The Active Intrusion Detection System more commonly known as Intrusion Prevention Systems is configured to respond to the attack in case of a security threat. Whenever there is an attack the Active Intrusion Detection System automatically takes action to deal with the attack in real time.

#### 3.3.2 PASSIVE INTRUSION DETECTION SYSTEM

The Passive Intrusion Detection System on the other hand does not take any action in case of an attack but only alerts the user of the vulnerabilities to the system.

## 4. ARTIFICIAL INTELLIGENCE IN INTRUSION DETECTION SYSTEMS

As mentioned in Section 3.2, Intrusion Detection Systems adopts mainly two strategies for detection of threat, the Signature based detection technique and the Anomaly based detection technique. As the security threats in the Signature based detection techniques are already predefined, they can be termed static. But for Anomaly based detection techniques are based heavily on Artificial Intelligence for fighting against the security threats. Some of them are Statistical based, Operational or threshold metric model, Markov Process or Marker Model, Statistical Moments or mean and standard deviation model, Univariate Model, Multivariate Model, Time series Model, Cognition based, Finite State Machine Model, Description script Model, Adept System Model, Machine Learning based, Bayesian Model, Genetic Algorithm model, Neural Network Model, Fuzzy Logic Model, Outlier Detection Model, Computer Immunology based and User Intention based Model [12, 20].

## 5. IDS BASED ON FUZZY LOGIC

Fuzzy Logic is a reasoning technique in which the reasoning is not precise and fixed but rather is an approximate value. Fuzzy Logic can therefore be aptly applied to Intrusion Detection Systems [8, 9, 32, 15, 34] to decide about suspected behavior when there is no clear distinction between anomalous and normal behavior in the traffic pattern. In addition to this, Fuzzy Logic greatly reduces the false positive alarm rate in Intrusion Detection Systems [11, 17].

The Fuzzy Logic uses the fuzzy variable along with the membership function to determine whether a particular rule is applicable to classify the condition as an anomaly or not.

### 5.1 APPLICATION OF FUZZY LOGIC TO IDS

The application of Fuzzy Logic to Intrusion Detection System has the following form

*If* **condition** *then* **consequence**

where,
  ***Condition*** is a fuzzy variable.
  ***Consequence*** is the fuzzy set.

Let us consider a typical scenario,

*If* number of packets with same destination address is **HIGH**
*then* pattern is unusual.

Now, to determine how many numbers of packets are considered in the category HIGH, the values of packets should be divided into some discrete sets known as fuzzy sets.
We consider a fuzzy space of three sets LOW, MEDIUM and HIGH, then

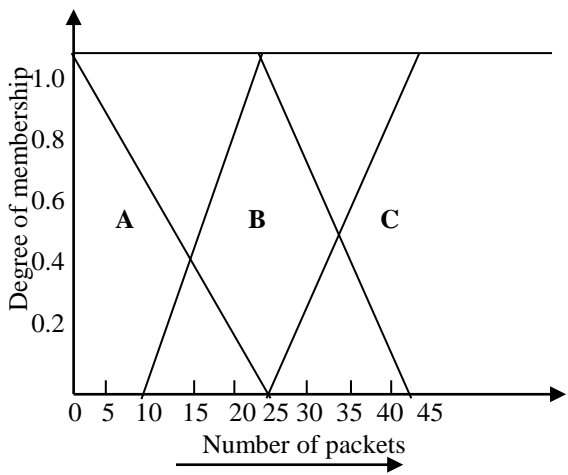

Figure 2. Fuzzy space used in Intrusion Detection

The region A depicts fuzzy set LOW, B depicts fuzzy set MEDIUM and C depicts fuzzy set HIGH. The x axis shows the values in the fuzzy set and the y axis shows the membership function. The number of packets is the fuzzy variable which is also known as fuzzy linguistics whereas the LOW, HIGH and MIDDLE depicts the values of the fuzzy variable.
So, after the intrusion detection system analyses the packets and finds the packets with the same destination number to be 15, then this condition will be regarded as LOW for a degree 0.4, but for a degree 0.6 this will be regarded as HIGH.

So in the Intrusion Detection System using Fuzzy Logic, this can be written as,

IF NumberOfPackets = HIGH
THEN AbortConnection.

Here, NumberOfPackets is the fuzzy variable and HIGH is the fuzzy set. So, depending upon the number of packets in the incoming traffic and the membership function, the value of HIGH is determined and accordingly the Intrusion Detection System will decide whether to abort the connection or not.

### 5.2 RELATED WORKS IN FUZZY LOGIC BASED IDS

Many researchers have made significant contribution towards Intrusion Detection System using Fuzzy Logic.
Chavan *et al*. [8] used Fuzzy Inference System combined with Artificial Neural Networks for real time traffic analysis. A signature pattern database was then constructed using protocol analysis and neuro fuzzy learning techniques.
In [9], Abraham *et al.* used three Fuzzy rule based classifiers for Intrusion Detection System and based on the rules, a soft computing based Intrusion Detection System was modeled.
Dickerson and Dickerson [32] proposed Fuzzy Intrusion Recognition Engine (FIRE). FIRE used Fuzzy Logic and data mining technique to produce fuzzy sets based on the input traffic data. The rules are then produced by fuzzy sets to detect the security threats. The limitation of FIRE as mentioned by the authors was its labor intensive rule generation process.
Researchers Tajbakhsh *et al.* [15] proposed a system to classify normal and anomalous attacks on the basis of compatibility threshold. Here the researchers used association based classification to classify the network traffic data.
Barbara *et al.* [34] proposed Audit Data Analysis and Mining (ADAM) which was a real time anomaly detection system. Here, suspicious events were classified as false alarms or real attacks by a module using association rules along with data mining techniques and classification.

## 6. IDS BASED ON GENETIC ALGORITHMS

Genetic Algorithms are biologically inspired search heuristics that employs evolutionary algorithm techniques like crossover, inheritance, mutation, selection etc. So, genetic algorithms are capable of deriving classification rules [1] and selecting optimal parameters for detection process [2]. Therefore Genetic Algorithms can be used to derive classification rules and apply them in Intrusion Detection System [1, 2, 13, 27, 28].

## 6.1 PARAMETERS USED IN GENETIC ALGORITHMS

1. **Fitness Function :**
   The fitness function evaluates the quality of a particular solution. The fitness function is used to select the best solution among all the solutions in the population. The fitness function should be an optimized value [43].

2. **Selection:**
   Selection is the process of choosing solution with better fitness function than their counterparts. In the selection phase the solutions having better fitness function over other solutions are selected and the rest are discarded.

3. **Crossover:**
   Crossover is the phase in which two solutions exchange one of their characteristics with the other in the pair at a randomly selected crossover point, where the crossover probability is between 0.6 and 0.9[25]. The solutions selected for crossover operation should be different [44].

4. **Mutation:**
   Mutation is a process by which some random bits in a solution are changed. This is done mainly to maintain the genetic diversity of the solutions.

## 6.2 APPLICATION OF GENETIC ALGORITHM TO IDS

The application of Genetic Algorithm to the network data consist primarily of the following steps:

1. The Intrusion Detection System collects the information about the traffic passing through a particular network.

2. The Intrusion Detection System then applies Genetic Algorithms which is trained with the classification rules learned from the information collected from the network analysis done by the Intrusion Detection System.

3. The Intrusion Detection System then uses the set of rules to classify the incoming traffic as anomalous or normal based on their pattern.

For application of genetic algorithm to Intrusion Detection System, we must represent the solution of our problem as a chromosome or genome. At the beginning a randomly generated population is selected. Three genetic operators, selection, mutation and crossover are applied to each generation and the best solution is found out. With passing generations newer generations evolve with better qualities than their previous generations [7, 23, 22, 35].

## 6.2 REPRESENTING DATA IN INTRUSION DETECTION SYSTEM

In Intrusion Detection System, the genes can be represented as different data types like integer, float and bytes [7].The genetic algorithm can be aptly applied to differentiate between anomalous network connections from normal ones The rule stored in the rule base is of the following form [4]

*If <condition>  then <action>*

The rule is basically a set of *if clause* where the condition is matched with the rules already stored in the IDS. The rule could specify the details of the packet like IP address, port number and protocol. If the incoming packet matches with any of the rules set in the IDS which have been classified as a threat, the IDS immediately takes action which may include alarming the system, stopping the connection or logging off the system [1,7].

Let us take a typical example,

The data representations of different genes are:

| Attribute | Number of Genes |
|---|---|
| Source Address | 4 |
| Destination Address | 4 |
| Packet Size | 1 |
| Port Number | 1 |

TABLE 1. Representation of different attributes

Here, the source address and destination address has four parts which is in the form of *w.x.y.z*. The packet size and the port number can be represented using one gene.

If the Intrusion Detection System has the rule set,

**If {   source address = "117.11.8.24";
    destination address = "117.15.9.25";
    packet size = "250";
    port number = 5005
    }**

**then {abort connection};**

In case of the above rule the Intrusion Detection System would analyze the incoming packet. If the packet has a source address of "117.11.8.24", destination address of "117.15.9.25", the packet size is "250" and the port number mentioned in the packet is "5005", then according to the rules, the packet will be deemed as a threat and accordingly the Intrusion Detection System will abort the connection.

The Intrusion Detection System using the genetic algorithm would convert the data into a chromosome of the following form

{117, 11, 8, 24, 117, 11, 9, 25, 250, 5005}

But to make a more general rule, wild card entries which are -1 are used. To generalize a source address of the form 117.11.*.*, the rule * will be changed with -1, so the chromosome will be,

{117, 11, -1, -1, 117, 15, 9, 25, 250, 5005}

### 6.3 Related works in genetic algorithm based IDS

A lot of work has been done on Intrusion Detection Systems using Genetic Algorithms.

Li [1] used genetic algorithms to identify anomalous network behaviors by considering both temporal and spatial information of the network connection during the encoding phase

Bridges and Vaughn [2] combined Genetic Algorithm and Fuzzy data mining techniques to detect network anomalies and misuses. Here the researchers applied Genetic Algorithm to find the best possible feature of fuzzy function and select the most significant network feature.

Lu and Traore [27] used Genetic programming to derive classification rules with traffic data of the network.

Crosbie and Spafford [28] used multiple agent technology and genetic programming to detect anomalies in the network. One limitation of the model was that when the agents were not initialized properly, the training process took a long time.

Xia *et al.* [13] used a combination of Genetic Algorithms and information theory to detect anomalous behaviors in the network theory. The approach used information theory to filter the traffic data which reduced the complexity.

### 7. Conclusion

In this paper an overview of the Intrusion Detection System is given. The approach to use Genetic Algorithms and Fuzzy Logic in Intrusion Detection System was also discussed. With the increasing use of the internet, the security threats have multiplied many folds. Along with all other conventional method, Intrusion Detection System have come a long way in the fight against security vulnerabilities. The use of Genetic Algorithms in Intrusion Detection System is particularly useful as it considers both temporal and spatial information of the network connections [1]. Moreover the use of fuzzy logic can help in detecting anomalies which cannot be discreetly deemed as normal or anomalous.

**References**

[1] Wei Li, "Using Genetic Algorithm for Network Intrusion Detection", *Proceedings of the United States Department of Energy Cyber Security Group*, 2004, pp. 1-8.

[2] Susan M Bridges, Rayford B. Vaughn, "Fuzzy data mining and genetic algorithms applied to intrusion detection.", *Proceedings of the National Information Systems Security Conference*; 2000.p. 13–31.

[3] Soumyendu Das, Subhendu Das, Bijoy Bandyopadhyay and Sugata Sanyal, "Steganography and Steganalysis: Different Approaches", *International Journal of Computers, Information Technology and Engineering (IJCITAE), Vol. 2*, No 1, June, 2008, Serial Publications.

[4] Sara Matzner, Chris Sinclair, Lyn Pierce, "An Application of Machine Learning to Network Intrusion Detection", *Proceedings of the 15th Annual Computer Security Applications Conference*, December 1999, Phoenix, AZ, pp. 331-371.

[5] Vipul Goyal, Virendra Kumar, Mayank Singh, Ajith Abraham and Sugata Sanyal, "A New Protocol to Counter Online Dictionary Attacks", *Computers and Security , Volume 25*, Issue 2 , Elsevier Science, March, 2006, pp. 114-120.

[6] Simon Liu, Mark Silverman, "A practical guide to biometric security technology", *IT Professional 3* (1) (2001), pp. 27–32.

[7] Ren Hui Gong, Mohammad Zulkernine, and Purang Abolmaesumi, "A software implementation of a genetic algorithm based approach to network intrusion detection", *The sixth International Conference on Software Engineering, Artificial Intelligence, Networking and Parallel/Distributed Computing, 2005 and the First ACIS International Workshop on Self-Assembling Wireless Networks(SNPD/SAWN '05),* pages 246 – 253.

[8] Sampada Chavan, Khusbu Shah, Neha Dave, Sanghamitra Mukherjee, Ajith Abraham and Sugata Sanyal, "Adaptive Neuro-Fuzzy Intrusion Detection Systems", *IEEE International Conference on Information Technology: Coding and Computing,2004.(ITCC '04), Proceedings of ITCC 2004, Vol. 1*, April, 2004, Las Vegas, Nevada, USA pp. 70-74

[9] Ajith Abraham, Ravi Jain, Sugata Sanyal and Sang Yong Han, "SCIDS: A Soft Computing Intrusion Detection System", *6th International Workshop on Distributed Computing (IWDC-2004),* Springer Verlag, Germany, Lecture Notes in Computer Science, Vol. 3326. ISBN: 3-540-24076-4, 2004, pp. 252-257.

[10] Jing T. Yao, Song L. Zhao, and Larry. V. Saxton, "A study on fuzzy intrusion detection", *Proceedings of SPIE: Data Mining, Intrusion Detection, Information Assurance, And Data Networks Security, volume 5812*, 2005 pp. 23–30.

[11] Vipul Goyal, Ajith Abraham, Sugata Sanyal and Sang Yong Han, "The N/R One Time Password System." *Information Assurance and Security Track*

*(IAS'05), IEEE International Conference on Information Technology: Coding and Computing (ITCC'05)*, USA, April, 2005, IEEE Computer Society, pp. 733-738.

[12] V. Jyothsna, Rama V. V. Prasad and Munivara K. Prasad, "A Review of Anomaly based Intrusion Detection Systems", *International Journal of Computer Applications 28* (7), August 2011. Foundation of Computer Science, New York, USA, pp. 26-35.

[13] Tao Xia, Guangzhi Qu, Salim Hariri and Mazin Yousif, "An efficient Network Detection Method Based on Information Theory and Genetic Algorithm", *Proceedings of the 24th IEEE International Performance Computing and Communications Conference*, Phoenix, Arizona, USA, 2005.

[14] Rangarajan A. Vasudevan, Ajith Abraham, Sugata Sanyal and Dharma P. Agrawal, "Jigsaw-based Secure Data Transfer over Computer Networks", *IEEE International Conference on Information Technology: Coding and Computing, 2004. (ITCC '04), Proceedings of ITCC 2004, Vol. 1*, April, 2004, Las Vegas, Nevada, USA, pp. 2-6.

[15] Arman Tajbakhsh, Mohammad Rahmati, Abdolreza Mirzaei, "Intrusion detection using fuzzy association rules", *Applied Soft Computing 9*(2):pp. 462-469.

[16] Ayu Tiwari, Sudip Sanyal, Ajith Abraham and Sugata Sanyal, "A Multifactor Security Protocol For Wireless Payment-Secure Web Authentication using Mobile Devices", *IADIS International Conference, Applied Computing 2007*, Salamanca, Spain. 17-20 February, 2007.

[17] Jonathan Gomez, Dipankar Dasgupta, "Evolving Fuzzy Classifiers for Intrusion Detection", *Proceeding of the 2002 IEEE, Workshop on Information Assurance*, United States Military Academy, June 2001, West Point, NY.

[18] Rangarajan A. Vasudevan and Sugata Sanyal, "A Novel Multipath Approach to Security in Mobile and Ad Hoc Networks (MANETs)", *Proceedings of International Conference on Computers and Devices for Communication (CODEC'04)*, Kolkata, India, 2004,pp.CAN_0412_CO_F_1to AN_0412_CO_F_4.

[19] R.G.Bace, "Intrusion Detection", *Macmillan Technical Publishing*, Indianapolis, USA, 2000.

[20] P. Garcia Teodoro, J Diaz Verdejo, G. Marcia Fernandez, E. Vazquez, "Anomaly-based network intrusion detection: Techniques, systems and challenges", *Computer and Security, 28*, 2009, pp. 18–28.

[21] Dhaval Gada, Rajat Gogri, Punit Rathod, Zalak Dedhia, Nirali Mody, Sugata Sanyal, "Security scheme for distributed DoS in mobile ad hoc networks"*, Arxiv preprint arXiv:1005.0109*,May 2010.

[22] Marczyk, A. "Genetic Algorithms and Evolutionary Computation." The Talk, Origins Archive. 23 Apr. 2004.7 Oct 2006 http://www.talkorigins.org /faqs/ genalg/genalg.html.

[23] K. M Faraoun, A. Boukelif. "Genetic Programming Approach for Multi-Category Pattern Classification Applied to Network Intrusions Detection.", *International Journal Of Computational Intelligence , Vol. 3*, No. 1, 2006 pp. 79-90

[24] Animesh Kr Trivedi, Rishi Kapoor, Rajan Arora, Sudip Sanyal and Sugata Sanyal, "RISM - Reputation Based Intrusion Detection System for Mobile Ad hoc Networks", *Third International Conference on Computers and Devices for Communications, CODEC-06, Institute of Radio Physics and Electronics*, University of Calcutta, December 18-20, 2006, Kolkata, India, pp. 234-

[25] IMSL C Numerical Stat Library, "Genetic Algorithms–An Overview." http://www.roguewave.com/portals/ 0/products/imsl-numerical-libraries/clibrary/docs/7.0/html/cstat/default.htm?turl=geneticalgorithmsanoverview.htm

[26] Sandipan Dey, Ajith Abraham and Sugata Sanyal , "An LSB Data Hiding Technique Using Prime Numbers", *Third International Symposium on Information Assurance and Security*, August 29-31, 2007, Manchester, United Kingdom, IEEE Computer Society press, USA, ISBN 0-7695-2876-7, pp. 101-106, 2007.

[27] Wei Lu and Issa Traore, "Detecting new forms of network intrusion using genetic programming", *Computational Intelligence Vol.20*, Issue 3, August 2004, pages 475-494.

[28] Mark Crosbie, and Eugene H. Spafford. "Applying Genetic Programming to Intrusion Detection". In *Proceeding of 1995 AAAI Fall Symposium on Genetic Programming*, Cambridge, Massachusetts, 1995, pp. 1-8.

[29] B.V. Dasarathy, "Intrusion Detection", *Information Fusion, Vol.4*, No.4, pp.243-245, 2003.

[30] J. Allen, A. Christie, and W. Fithen, "State Of the Practice of Intrusion Detection Technologies", *Technical Report, CMU/SEI-99-TR-028*, 2000.

[31] Dhaval Gada, Rajat Gogri, Punit Rathod, Zalak Dedhia, Nirali Mody, Sugata Sanyal and Ajith Abraham, "A Distributed Security Scheme for Ad Hoc Networks", *ACM Crossroads, Special Issue on Computer Security. Volume 11*, No. 1, September, 2004, pp. 1-17.

[32] John E. Dickerson, Julie A. Dickerson, "Fuzzy Network Profiling for Intrusion Detection". *Proceedings of 19th International Conference of the North American Fuzzy Information Processing Society*, Atlanta, 2000, pp.301-306.

[33] Bhavyesh Divecha, Ajith Abraham, Crina Grosan and Sugata Sanyal, "Analysis of Dynamic Source Routing and Destination-Sequenced Distance-Vector Protocols for Different Mobility models"*, First Asia International Conference on Modelling and Simulation, AMS2007*. 27-30 March, 2007, Phuket, Thailand.

[34] Daniel Barbará., Julia Couto, Sushil Jajodia, Ningning Wu. "ADAM: a testbed for exploring the

use of data mining in intrusion detection", *ACM SIGMOD 30(Special)*, 2001, pp. 15-24.

[35] Dong Song, "A linear genetic programming approach to intrusion detection", Master's thesis, Dalhousie University, March 2003.

[36] John Bellardo and Stefan Savage, "802.11 denial-of-service attacks: Real vulnerabilities and practical solutions". *In Proceedings of the USENIX Security Symposium*, 2003, pp. 15- 28

[37] Christoph L. Schuba, Ivan Krsul, Markus G. Kuhn, Eugene H. Spafford, Aurobindo Sundaram, and Diego Zamboni. "Analysis of a denial of service attack on TCP", In *Proceedings of the 1997 IEEE Symposium on Security and Privacy* May 1997.

[38] A. B. Moreno and A. S´anchez. "GavabDB: a 3D face database". *In Workshop on Biometrics on the Internet, Vigo*, March 2004, pp.77–85

[39] V. Bobor, "Efficient Intrusion Detection System Architecture Based on Neural Networks and Genetic Algorithms", *Department of Computer and Systems Sciences, Stockholm University / Royal Institute of Technology, KTH/DSV*, 2006.

[40] Vipul Goyal, Virendra Kumar, Mayank Singh, Ajith Abraham and Sugata Sanyal, "CompChall: Addressing Password Guessing Attacks", *Information Assurance and Security Track (IAS'05), IEEE International Conference on Information Technology: Coding and Computing (ITCC'05),* USA. April 2005, IEEE Computer Society, pp. 739-744.

[41] Chi Ho Tsang, Sam Kwong, and Hanli Wang, "Genetic-fuzzy rule mining approach and evaluation of feature selection techniques for anomaly intrusion detection." *Pattern Recognition, 40*(9), 2007, pp. 2373–2391.

[42] M. Saniee Abadeh, J. Habibi, and C. Lucas, "Intrusion detection using a fuzzy genetics-based learning algorithm." *Journal of Network and Computer Applications, 30*(1), 2007, pp. 414–428.

[43] Adel Nadjaran Toosi, Mohsen Kahani, "A new approach to intrusion detection based on an evolutionary soft computing model using neuro-fuzzy classifiers", *Computer Communications, 30*(10), 2007, pp. 2201–2212.

[44] Mohammad Saniee Abadeh, Jafar Habibi, Zeynab Barzegar, Muna Sergi, "A parallel genetic local search algorithm for intrusion detection in computer Networks", *Engineering Applications of Artificial Intelligence, 20*(8), 2007, pp.1058–1069.